\documentclass[12pt]{article}
\usepackage[english,german,french,polish]{babel}
\usepackage[T1]{fontenc}
\usepackage{amsfonts}

\begin{document}

\selectlanguage{english}

\baselineskip 0.73cm
\topmargin -0.4in
\oddsidemargin -0.1in

\let\ni=\noindent

\renewcommand{\thefootnote}{\fnsymbol{footnote}}

\newcommand{\SM}{Standard Model }

\newcommand{\SMo}{Standard-Model }

\pagestyle {plain}

\setcounter{page}{1}



~~~~~~
\pagestyle{empty}

\begin{flushright}
\end{flushright}

\vspace{0.4cm}

{\large\centerline{\bf A new weak-isospin triplet of scalars and "electroweak portal"}}

{\large\centerline{\bf to hidden sector of the Universe}}

\vspace{0.5cm}

{\centerline {\sc Wojciech Kr\'{o}likowski}}

\vspace{0.3cm}

{\centerline {\it Institute of Theoretical Physics, University of Warsaw}}

{\centerline {\it Ho\.{z}a 69, 00--681 Warszawa, ~Poland}}

\vspace{0.6cm}

{\centerline{\bf Abstract}}

\vspace{0.2cm}

\begin{small}


A new weak-isospin triplet of scalar bosons is introduced to the \SM in order to ensure the electroweak symmetry
for a conjectured new weak coupling between a hidden sector of the Universe, responsible for cold dark matter,
and the Standard-Model sector (involving now the new triplet). The considered coupling may be called  
"electroweak portal"~to the structure of hidden sector. After spontaneously breaking the electroweak symmetry
by the \SMo Higgs mechanism, it includes the so called~"photonic portal"~introduced by us previously. 
 
\vspace{0.6cm}

\ni PACS numbers: 12.60.-i , 12.60.Cn , 12.60.Fr 

\vspace{0.3cm}

\ni November 2012

 
\end{small}

\vfill\eject

\pagestyle {plain}

\setcounter{page}{1}


Not long ago, we speculated that in the Universe --- beside the \SMo sector --- there is a hidden sector responsible for the cold dark matter, coupled weakly to the former through the so called "photonic portal"~[1,2] (in contrast to the popular "Higgs portal"~ usually considered [3]).

It was tentatively conjectured that such a sector consists of the following fields (all sterile with respect to the \SMo charges): spinor $\psi(x)$ ("sterino"), scalar $\varphi(x)$ ("steron") and mediating antisymmetric tensor $A_{\mu\,\nu}(x)$ of dimension one ("$\!A$ boson"). These fields are coupled weakly to the \SMo sector and to themselves through the gravity as well as through two interactions:

\vspace{-0.2cm}

\begin{equation}
-\frac{1}{2} \left(\sqrt{f\,} \sum_i \varphi_i W^{\mu \nu}_i + \sqrt{f'\,} \varphi B^{\mu \nu} \right)A_{\mu\,\nu} 
\end{equation}

\vspace{-0.1cm}

\ni and 

\vspace{-0.2cm}

\begin{equation}
-\frac{1}{2} \zeta  \sqrt{f\,} \bar\psi \sigma^{\mu\,\nu} \psi  A_{\mu \nu}
\end{equation}

\vspace{-0.1cm}

\ni conserving the electroweak symmetry. Here, $\sqrt{f\,} , \sqrt{f'\,}$ and $\zeta \sqrt{f\,}$ ($f>0$ and $f'>0$) are dimensionless coupling constants, $\varphi_i(x)\; (i=1,2,3)$ is a new weak-isospin triplet of scalar fields with zero weak hypercharge (extending the \SM): 

\vspace{-0.2cm}

\begin{equation}
\varphi_1 = \frac{1}{\sqrt{2}} (\varphi^+ + \varphi^-) \; , \;\varphi_2 = \frac{1}{\sqrt{2}} (\varphi^+ - \varphi^-) \;, \;\varphi_3 = \varphi^0 
\end{equation}

\vspace{-0.1cm}

\ni and, finally, $W^{\mu \nu}_i(x)$ and $B^{\mu \nu}(x)$ are Weinberg-Salam antisymmetric tensors of dimension two mediating electroweak interactions: 

\vspace{-0.2cm}

\begin{equation} 
W^{\mu \nu}_i= \partial^\mu W^{\nu}_i  - \partial^\nu W^{\mu}_i + g\sum_{j\,k} \varepsilon_{ijk} W^{\mu}_j 
W^{\nu}_k \,\;,\,\; B^{\mu \nu} = \partial^\mu B^\nu - \partial^\nu B^\mu \;,
\end{equation}

\vspace{-0.1cm}

\ni where $W^{\mu}_i(x)\;(i=1,2,3)$ and $B^\mu(x)$ denote the electroweak gauge fields. Beside the coupling (1) that we will call the "\,$\!$electroweak portal", the new scalar fields (3) can interact also through the conventional electroweak gauge coupling of the \SM (extended to include this scalar triplet as a kind of "matter"). After spontaneously breaking the electroweak symmetry by the \SMo Higgs mechanism involving the weak-isospin doublet of scalar fields, we get in Eq. (1):

\vspace{-0.3cm}

\begin{eqnarray} 
W^{\mu}_1\! = \frac{1}{\sqrt{2}}(W^{+ \mu} + W^{- \mu}) \;,\;W^{\mu}_2\! = \frac{1}{i\sqrt{2}}(W^{+ \mu}\!\! & \!\!-\!\! & \!\!W^{- \mu}) \;,\; W^{\mu}_3\! = W^{0 \mu}\! = \cos\theta_w Z^\mu - \sin\theta_w A^\mu\;, \nonumber \\
 B^{\mu} = \sin\theta_w Z^\mu\!\! &\!\!+\!\! & \cos\theta_w A^\mu\,.
\end{eqnarray}

\vspace{-0.1cm}

\ni In addition, we put then $<\!\!\varphi_3\!\!>_{\rm vac} = 0$ for the weak-isospin triplet $\varphi_i(x)\;(i=1,2,3)$, but assume that $\varphi(x) = {<\!\varphi\!>}_{\rm vac} + \,\varphi_{\rm ph}(x)$ with $<\!\!\varphi\!\!>_{\rm vac} \neq 0$ for the steron $\varphi(x)$. The charged scalars $\varphi^{\pm}(x)$ appear in this case beside the neutral $\varphi^0(x)$ as physical fields, not being eaten by $W_\mu^{\pm}(x)$ and $Z_\mu(x)$. The mass of $\varphi_i(x) \;(i = 1,2,3)$ can be produced as $m_{\vec{\varphi}} = \sqrt{Y_{\vec{\varphi}}}<\!\!h^0\!\!>_{\rm vac}$, where $h^0(x) = <\!\!h^0\!\!>_{\rm vac}+ h^0_{\rm ph}(x)$ denotes the neutral component of \SMo Higgs doublet, while $m_{\varphi} = \sqrt{\lambda_\varphi}<\!\!\varphi\!\!>_{\rm vac}$ and $M = \sqrt{\lambda_A}<\!\!\varphi\!\!>_{\rm vac}$ can be masses of $\varphi_{\rm ph}(x)$ and $A_{\mu \nu}(x)$, respectively. 

Inserting Eqs. (3), (4) and (5) into the interaction (1), we obtain the electroweak portal in the form

\vspace{-0.3cm} 

\begin{eqnarray} 
-\frac{1}{2}(\!\!\sqrt{\!f}\! &\!\!\!\sum_i\!\!\! &\!\varphi_iW_i^{\mu \nu}\!\! + \sqrt{\!f'}\varphi B^{\mu \nu}\!) \!A_{\mu \nu} \nonumber \\ 
& \!\!\!=\!\! &\!\!\! -\frac{1}{2}  \sqrt{\!f\,}  \varphi^- \left[\partial^\mu W^{+ \nu} -  \partial^\nu W^{+ \mu}- i g \left(W^{+\mu} W^\nu_3 - W^\mu_3 W^{+\nu} \right)\!\right] A_{\mu \nu} \nonumber \\ 
& \!\!\!\! &\!\!\! -\frac{1}{2}\sqrt{\!f\,} \varphi^+ \left[\partial^\mu W^{- \nu} -  \partial^\nu W^{-\mu} +  i g \left(W^{-\mu} W^\nu_3 - W^\mu_3W^{-\nu}\right)\!\right] A_{\mu \nu} \nonumber \\ 
& \!\!\!\! &\!\!\! -\frac{1}{2}\!\left[\!\sqrt{\!f^{(Z)\,}} \varphi^{(Z)} Z^{\mu \nu} \!\!+\! \sqrt{\!f^{(\gamma)\,}}\varphi^{(\gamma)}F^{\mu \nu}\!\!+\! \sqrt{\!f\,} ig \varphi_3\! \left(\!W^{+\mu}W^{-\nu} \!\!\!-\! W^{-\mu}W^{+\nu} \!\right)\!\right] \!A_{\mu \nu} , 
\end{eqnarray}

\ni where $W^\mu_3  = \cos\theta_w Z^\mu - \sin\theta_w A^\mu $ and

\vspace{-0.2cm}

\begin{equation} 
Z^{\mu \nu}  = \partial^\mu Z^{\nu} - \partial^\nu Z^{\mu} \;\;,\;\; F^{\mu \nu} = \partial^\mu A^{\nu} \!-\!  \partial^\nu A^{ \mu}\,,
\end{equation}

\vspace{-0.1cm}

\ni while 

\vspace{-0.3cm}

\begin{equation} 
f^{(Z)} \equiv f \cos^2\theta_w + f' \sin^2\theta_w \;\,,\;\, f^{(\gamma)} \equiv f \sin^2\theta_w + f' \cos^2\theta_w \,,
\end{equation}

\vspace{-0.1cm}

\ni and 

\vspace{-0.3cm}

\begin{eqnarray}
\sqrt{\!f^{(Z)\,}} \varphi^{(Z)} & \equiv & \sqrt{\!f} \cos\theta_w \varphi_3 + \sqrt{\!f' }\sin\theta_w \varphi \,, \nonumber \\
\sqrt{\!f^{(\gamma)\,}} \varphi^{(\gamma)} & \equiv & \!\!\!-\sqrt{\!f} \sin\theta_w \varphi_3 + \sqrt{\!f' }\cos\theta_w \varphi \,.
\end{eqnarray}

Note that in the hypothetic new weak interaction (6) there is embedded the extra coupling

\begin{equation}
-\frac{1}{2}\sqrt{\!f^{(\gamma)\,}}\varphi^{(\gamma)} F^{\mu \nu} A_{\mu \nu}
\end{equation}

\ni of electromagnetic field $F^{\mu \nu}(x) = \partial^\mu A_{\nu}(x) = \partial^\nu A_{\mu}(x)$, which we called the photonic portal [1,2]. Similarly, in the interaction there is involved the extra coupling 
 
\begin{equation}
-\frac{1}{2}\sqrt{\!f^{(Z)}}\, \varphi^{(Z)} Z^{\mu \nu} A_{\mu \nu}
\end{equation}

\ni of field $Z^{\mu \nu}(x) = \partial^\mu Z_{\nu}(x) - \partial^\nu Z_{\mu}(x)$.

Since after spontaneously breaking the electroweak symmetry we get still $<\!\varphi_3\!>_{\rm vac} = 0$ for the weak-isospin triplet $\varphi_i(x) \;\;(i=1, 2, 3)$, although  $\varphi(x) = {<\!\varphi\!>}_{\rm vac} + \,\varphi_{\rm ph}(x)$ with $<\!\!\varphi\!\!>_{\rm vac} \neq 0$ for the steron $\varphi(x)$, we infer from Eq. (9) that 

\vspace{-0.2cm}

\begin{equation} 
\varphi^{(\gamma)}(x) =<\!\!\varphi^{(\gamma)}\!\!>_{\rm vac} +\varphi^{(\gamma)}_{\rm \!ph}\;{\rm with}\;<\!\!\varphi^{(\gamma)}\!\!>_{\rm \!vac} = \sqrt{\!f'\!/\!f^{(\gamma)}} \cos\theta_w<\!\!\varphi\!>_{\rm \!vac} \neq 0
\end{equation}

\vspace{-0.2cm}

\ni and

\vspace{-0.1cm}

\begin{equation} 
\varphi^{(Z)}(x) = <\!\!\varphi^{(Z)} \!\!>_{\rm vac} + \varphi^{(Z)}_{\rm ph}\;{\rm with}\;<\!\!\varphi^{(Z)} \!\!>_{\rm vac} = \sqrt{\!f'\!/\!f^{(Z)}} \sin\theta_{w}<\!\!\varphi>_{\rm vac} \neq 0. 
\end{equation}

\vspace{0.1cm}

\ni Thus, the  interactions (10) and (11) are responsible for the virtual transitions

\begin{equation}
A^* \rightarrow \gamma \;\;{\rm and} \;\; A^* \rightarrow Z
\end{equation}

\vspace{-0.1cm}

\ni as well as for the virtual decays

\vspace{-0.2cm}

\begin{equation} 
\varphi^{(\gamma)}_{\rm ph} \rightarrow A^* \gamma \;\; {\rm and}\;\; \varphi^{( Z\!)}_{\rm ph} \rightarrow A^* Z
\end{equation}

\ni (an $A$ boson is described by the field $A_{\mu \nu}(x)$  of dimension one). So, we should observe the decays 

\begin{equation} 
\varphi^{(\gamma)}_{\rm ph} \rightarrow \gamma \gamma \;,\; 
\varphi^{(\gamma)}_{\rm ph} \rightarrow Z \gamma \;\;({\rm if}\;\; m_{\varphi^{(\!\gamma\!)}} > m_Z) 
\end{equation}

\ni and 

\vspace{-0.1cm}

\begin{equation}
\varphi^{(Z)}_{\rm ph} \rightarrow Z \gamma \;\;({\rm if}\;\; m_{\varphi^{(\!Z\!)}} > m_Z) \;,\;\varphi^{(Z)}_{\rm ph} \rightarrow ZZ \;\;({\rm if}\;\; m_{\varphi^{(\!Z\!)}} > 2m_Z) 
\end{equation}

\vspace{-0.2cm}

\ni as well as

\vspace{-0.1cm}

\begin{equation} 
\varphi^{(Z)}_{\rm ph} \rightarrow Z^*Z^* \rightarrow {\rm {two\;charged\!\!-\!\!lepton\; pairs}} \,,
\end{equation} 

\vspace{0.1cm}

\ni the last would proceed virtually through $Z^*Z^*$. Notice the existence in our case of tree-level channels $\gamma \gamma$ and $Z \gamma$ not appearing (on this level) in the \SMo Higgs case.

After some calculations based on the couplings (10) and (11) present in the electroweak portal (6) we obtain on the tree level the following widths for the decays $\varphi^{(\gamma)}_{\rm ph}\rightarrow \gamma \gamma$ and $\varphi^{(\gamma)}_{\rm ph} \rightarrow Z\gamma\;({\rm if}\;\; m_{\varphi^{(\!\gamma\!)}} > m_Z)$:  

\begin{equation} 
\Gamma(\varphi^{(\gamma)}_{\rm ph} \rightarrow \gamma\gamma)= \frac{f^{(\gamma)\,2}<\!\varphi^{(\gamma)}\!>\!_{\rm vac}^{\!2}}{128 \pi M^4}\, m^3_{\varphi^{(\gamma)}} \,,
\end{equation} 

\ni and

\begin{equation} 
\Gamma(\varphi^{(\gamma)}_{\rm ph} \rightarrow Z\gamma)= \frac{f^{(\gamma)}f^{(Z)}<\!\varphi^{(Z)}\!>\!_{\rm vac}^{\!2}}{64 \pi M^4}\, m^3_{\varphi^{(\gamma)}}\!\left(1- \frac{m^4_Z}{ m^4_{\varphi^{(\gamma)}} }\right)\,.
\end{equation} 

\ni Here, we used the Klein-Gordon equation for $A_{\mu\nu}(x)$, assuming that its mass scale $M^2$ dominates over the momentum transfers $\Box$ involved. Similarly, for the decay widths of  $\varphi^{(Z)}_{\rm ph} \rightarrow Z\gamma\;({\rm if}\;\; m_{\varphi^{(\!Z\!)}} > m_Z)$ and $\varphi^{(Z)}_{\rm ph} \rightarrow ZZ\;({\rm if}\;\; m_{\varphi^{(\!Z\!)}} > 2m_Z)$ we get 

\begin{equation} 
\Gamma(\varphi^{(Z)}_{\rm ph} \rightarrow Z\gamma) = \frac{ f^{(Z)} f^{(\gamma)}<\!\varphi^{(\gamma)}\!>\!_{\rm vac}^{\!2}}{ 64\pi M^4 }\, m^3_{\varphi^{(Z)}}\!\left(1- \frac{ m^4_Z }{ m^4_{\varphi^{(\gamma)}} }\right)  \,,
\end{equation} 

\ni and

\begin{equation} 
\Gamma(\varphi^{(Z)}_{\rm ph} \rightarrow ZZ) = \frac{ f^{(Z)\,2}<\!\varphi^{(Z)}\!>\!_{\rm vac}^{\!2}}{ 128 \pi M^4 } \,m^3_{\varphi^{(Z)}}\!\left( 1- 4\frac{ m^2_Z }{ m^2_{\varphi^{(Z)}} } + 6  
\frac{ m^4_Z }{ m^4_{\varphi^{(Z)}} }\right)\,.
\end{equation} 

\ni In particular, from Eqs. (19) and (20) it follows that

\begin{equation} 
\frac{ \Gamma(\varphi^{(\gamma)}_{\rm ph} \rightarrow Z\gamma) }{ \Gamma(\varphi^{(\gamma)}_{\rm ph} \rightarrow \gamma\gamma) } = 2\tan^2 \theta_w\!\left(1- \frac{m^4_Z}{ m^4_{\varphi^{(\gamma)}} }\right)\,,
\end{equation} 

\ni where Eqs. (12) and (13) are applied for $<\!\varphi^{(\gamma)}\!>\!_{\rm vac} $ and $<\!\varphi^{(Z)}\!>\!_{\rm vac} $, and so

\begin{equation} 
\frac{ f^{(Z)} <\!\varphi^{(Z)}\!>\!^2_{\rm vac} }{ f^{(\gamma)} <\!\varphi^{(\gamma)}\!>\!^2_{\rm vac} }= \tan^2 \theta_w \,.
\end{equation} 

\ni Here, $\tan^2 \theta_w = 0.301$ for the Weinberg angle and $m_Z = 91.2$ GeV for the mas of $Z$ boson.

If it happens that the masses $m_{\varphi^{(\!\gamma)}}\!$ and $m_{\varphi^{(\!Z)}}\!$ of physical scalar bosons $\varphi^{(\!\gamma)}_{\rm ph}\!$ and $\varphi^{(\!Z)}_{\rm ph}\!$ are nearly equal, then the channels $\varphi^{(\!\gamma)}_{\rm ph}\!\rightarrow \!\gamma\gamma$ and $\!\varphi^{(\!Z)}_{\rm ph}\! \rightarrow \!Z^*\!Z^*\!\rightarrow {\rm two\;charged\!\!-\!\!lepton\;pairs}$ ought to be distinguished from decay products of the neutral scalar boson of mass 125 GeV found recently at CERN in the channels $\rightarrow \gamma \gamma$ and $\rightarrow Z^*Z^* \rightarrow {\rm two\;charged\!\!-\!\!lepton\;pairs}$ [4], when this is really the \SMo Higgs boson as is expected (for a generic discussion on Higgs-boson imposters {\it cf.} Ref. [5]). 
 
\vfill\eject 

\vspace{2.0cm}

{\centerline{\bf References}}

\vspace{0.4cm}

\baselineskip 0.73cm 

{\everypar={\hangindent=0.65truecm}
\parindent=0pt\frenchspacing

{\everypar={\hangindent=0.65truecm}
\parindent=0pt\frenchspacing

~[1]~W.~Kr\'{o}likowski, {\it Acta Phys. Polon.} {\bf B 39}, 1881 (2008); {\it ibid.} {\bf B 40}, 111 (2009);  arXiv: 0803.2977v2 [{\tt hep--ph}].

\vspace{0.2cm}

~[2]~W.~Kr\'{o}likowski, {\it Acta Phys. Polon.} {\bf B 40}, 2767 (2009).

\vspace{0.2cm}

~[3]~J. March-Russell, S.M. West, D. Cumberbath and D.~Hooper, {\it J. High Energy Phys.} {\bf 0807}, 058 (2008).  

\vspace{0.2cm}

~[4]~J.~Incandela, the CMS Collaboration, talk given at CERN on July 4, 2012; F.Gianotti,  the ATLAS Collaboration, talk given at CERN on July 4, 2012.

\vspace{0.2cm}

~[5]~I.~Low, J.~Lykken and G. Shaughnessy, arXiv: 1207.1093 [{\tt hep--ph}].

\vfill\eject

\end{document}